\documentclass[twocolumn,10pt]{asme2e}

\usepackage{graphicx}
\graphicspath{{images/}}
\DeclareGraphicsExtensions{.pdf,.jpg,.jpeg,.png}
\usepackage{color}
\usepackage[caption=false,font=footnotesize]{subfig}

\usepackage{amsmath}
\usepackage{listings}

\begin{document}
\title{Dynamic Simulation of Soft Heterogeneous Objects}

\author{Jonathan Hiller
    \affiliation{
	Sibley School of Mechanical and Aerospace Engineering\\
	Cornell University\\
	Ithaca, New York 14850\\
    Email: jdh74@cornell.edu
    }	
}
\author{Hod Lipson
    \affiliation{
	Sibley School of Mechanical and Aerospace Engineering\\
	Computing and Information Science
	Cornell University\\
	Ithaca, New York 14850\\
    Email: hod.lipson@cornell.edu
    }	
}

\maketitle

\begin{abstract}
\noindent
{\bf Background:} This paper describes a 2D and 3D simulation engine that {\it quantitatively} models the statics, dynamics, and non-linear deformation of heterogeneous soft bodies in a computationally efficient manner. There is a large body of work simulating compliant mechanisms. These normally assume small deformations with homogeneous material properties actuated with external forces. There is also a large body of research on physically-based deformable objects for applications in computer graphics with the purpose of generating realistic appearances at the expense of accuracy. 

\noindent
{\bf Method of Approach:} Here we present a simulation framework in which an object may be composed of any number of interspersed materials with varying properties (stiffness, density, etc.) to enable true heterogeneous multi-material simulation. Collisions are handled to prevent self-penetration due to large deformation, which also allows multiple bodies to interact. A volumetric actuation method is implemented to impart motion to the structures which opens the door to the design of novel structures and mechanisms. 

\noindent
{\bf Results:} The simulator was implemented efficiently such that objects with thousands of degrees of freedom can be simulated at suitable framerates for user interaction using a single thread of a typical desktop computer. The code is written in platform agnostic C++ and is fully open source.

\noindent
{\bf Conclusions:} This research opens the door to the dynamic simulation of freeform 3D multi-material mechanisms and objects in a manner suitable for design automation.
\end{abstract}


\section{Introduction}
Scientific physics simulators are traditionally used to model small deformations of homogeneous linear-elastic materials. Recently, multi-material additive manufacturing methods have been developed that fabricate heterogeneous objects out of two or more materials. The properties of these co-fabricated materials can range from rigid plastics and metals to very soft rubber with linear deformation greater than 200\% \cite{Objet2011}. The inclusion of these soft, rubbery materials necessitates the consideration of large, non-linear geometric deformations to accurately predict the physical behavior. Because hard and soft materials can be internally combined and patterned in 3D with very few constraints, a new paradigm of physics simulation becomes necessary to efficiently predict the combined material properties and dynamic behavior.  

\begin{figure}[!t]
\centering
\subfloat[]{
\includegraphics[width=1.5in]{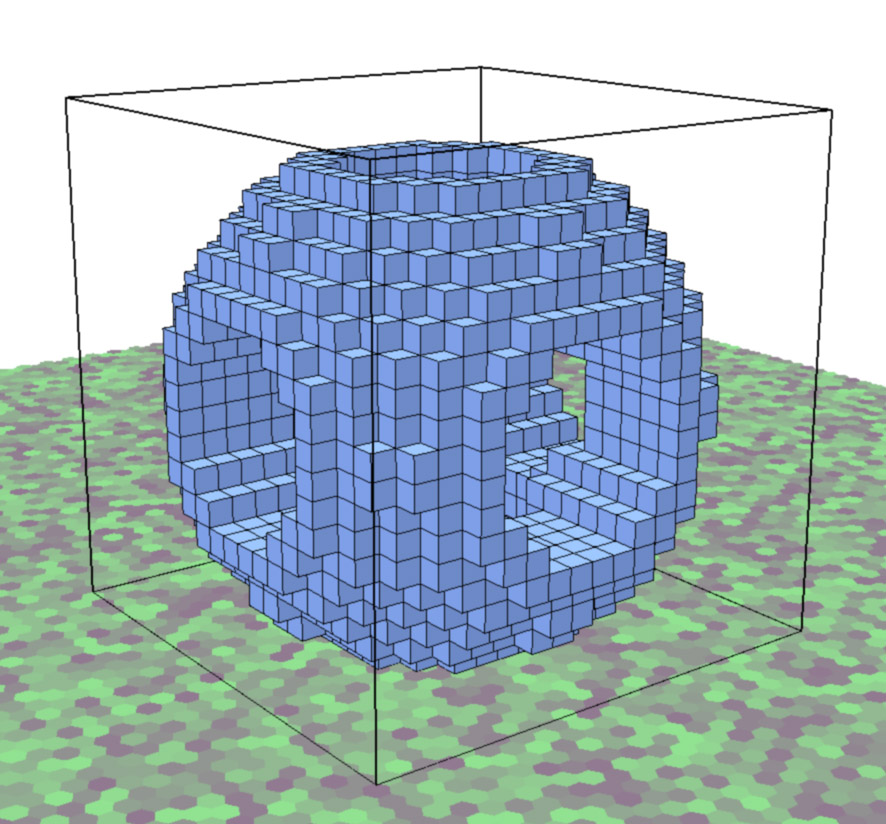}
\label{_Ball_falling}}
\subfloat[]{
\includegraphics[width=1.5in]{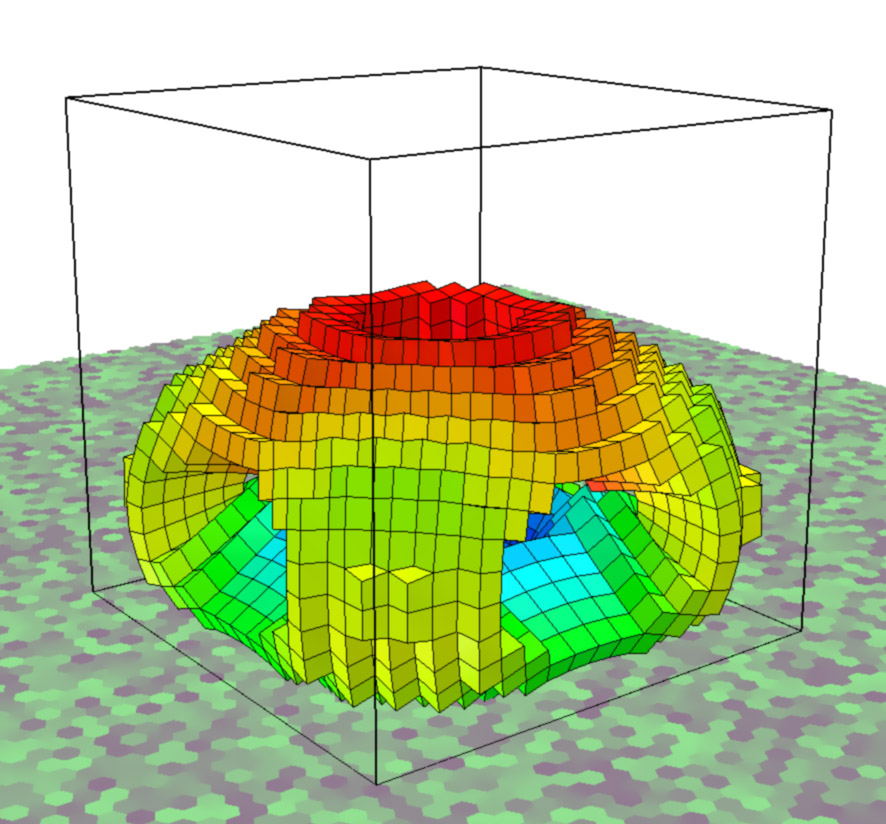}
\label{_Ball_bouncing}}
\caption{Two frames of a hollow ball bouncing under gravity illustrate the undeformed ball just before impact (a) and the highly deformed ball the moment of highest deformation (b).}
\label{Ball}
\end{figure}

There are many established methods and implementations for simulating deformable soft bodies \cite{Nealen2006}. Considering the large deformations and relatively low stiffness of the materials involved, the physically-based dynamics are often significant and must be modeled (Figure \ref{Ball}). Much of the development in simulating soft bodies has been driven by the computer graphics community. Many of the well established physics engines provide support for dynamic deformable bodies, whether 1D rope, 2D cloth, or 3D "jello" \cite{PhysX2011, Bullet2011}. 

The goal of these simulations is generally to create realistic visual effects in real time at the expense of accuracy \cite{Teschner2004, Eberhardt1996, Muller2002, Faloutsos1997}. For instance, lattice shape matching \cite{Rivers2007} creates visually appealing rubber effects very efficiently and is unconditionally stable. However, the underlying methods are geometrically based, which limits their direct application to quantitative engineering analysis problems. Other simulators are derived from more physically based principles \cite{James1999}, but their performance at predicting real-world behaviors is unverified. Deformable body simulators have also been developed specifically for real-time surgery simulation \cite{Cotin1999, Wu2001}. These simulators address challenges such as modifying the geometry dynamically to simulate incisions, but due to the variance of biological materials it is also difficult to verify quantitative accuracy.

There is also a large body of work regarding the simulation of compliant mechanisms and design thereof \cite{Lipson2006, Pederson2001}. Existing efforts focus on small displacements \cite{ Nishiwaki1998} or discrete thin beam members that can flex significantly \cite{saxena2001, howell1996}. In the simulation framework presented here, an entire freeform 3D shape can be non-linearly deformed, leading to many novel possibilities for soft mechanisms that cannot be simulated efficiently using current techniques.

\subsection{Finite Element vs. Mass-Spring methods}
Finite element analysis (FEA) is a well established method of simulating the mechanical behavior of objects. Advantages include the ability to solve a system with irregularly spaced discretized mesh elements. A stiffness matrix is composed containing information about the connectivity of the entire mesh and the local material properties at each node. However, this system can only be efficiently solved if the underlying equations are linear. Thus, deformations that change the geometry significantly require periodic re-meshing \cite{Wu2001}. Other non-linearities such as friction and advanced material models require additional levels of iteration to solve.

Mass-spring methods are widely used for deformable bodies, especially in dynamic simulations for computer graphics \cite{Nealen2006}. Advantages include relative simplicity and handling large deformations and other non-linearities with ease. An object is decomposed into discrete point masses connected by springs. Thus, the entire system forms a system of ordinary differential equations (ODEs) that can be integrated directly to solve for the behavior of the system. This makes these particle-based physical simulations very computationally efficient at the expense of accuracy.

\subsection{Freeform Mesh vs. Voxels on a Lattice}
There are a number of tradeoffs associated with choosing either a freeform mesh or a lattice of voxels to dynamically simulate a heterogeneous object. Both FEA methods and many existing gaming physics simulators use a freeform mesh to discretize a 3D object for simulation. By allowing the vertices to lie at any position within the object, there is greater control over the local detail of the simulation. Specifically, this allows objects to be meshed based on the desired accuracy in a given region or dynamically re-meshed based on the current regions of interest in a deformed shape \cite{Nesme2006}. Care must be taken when forming the mesh such that the aspect ratio of each element does not vary significantly in order to preserve accuracy. However the advantages of freeform meshes quickly diminish as materials of different stiffnesses and properties are interspersed within the object. This constrains the mesh generation process and can potentially create very large and inefficient meshes, such as the case of a dithering between two materials.

\begin{figure}[!t]
\centering
\subfloat[]{
\includegraphics[width=2.2in]{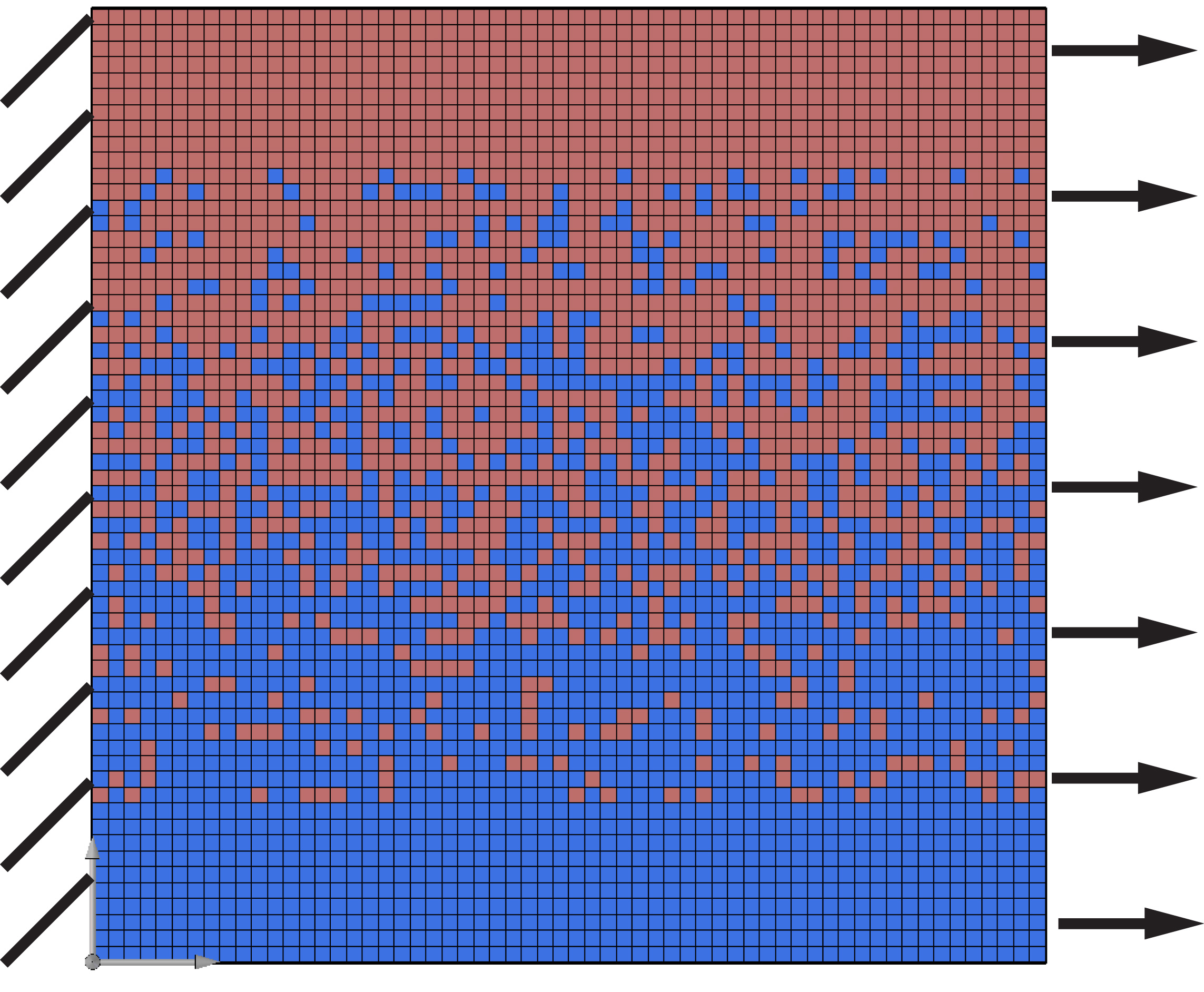}
\label{_Grad_Undeformed}}\
\subfloat[]{
\includegraphics[width=1.5in]{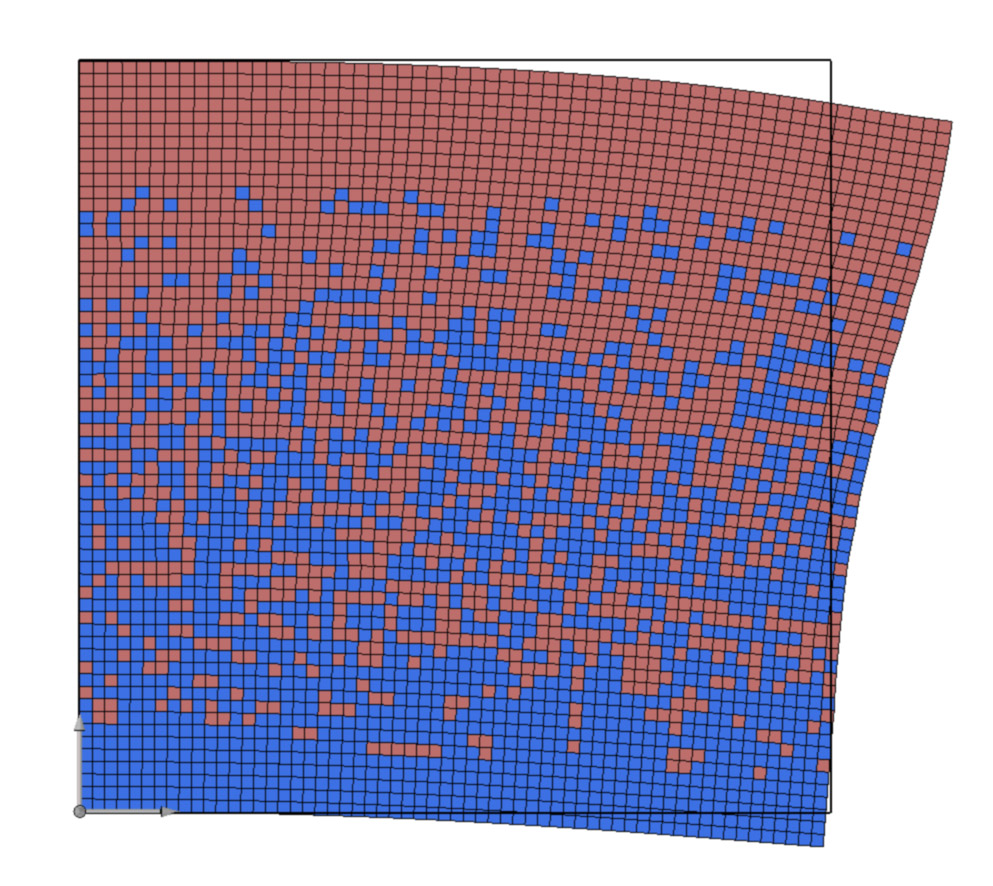}
\label{_Grad_Deformed}}
\subfloat[]{
\includegraphics[width=1.5in]{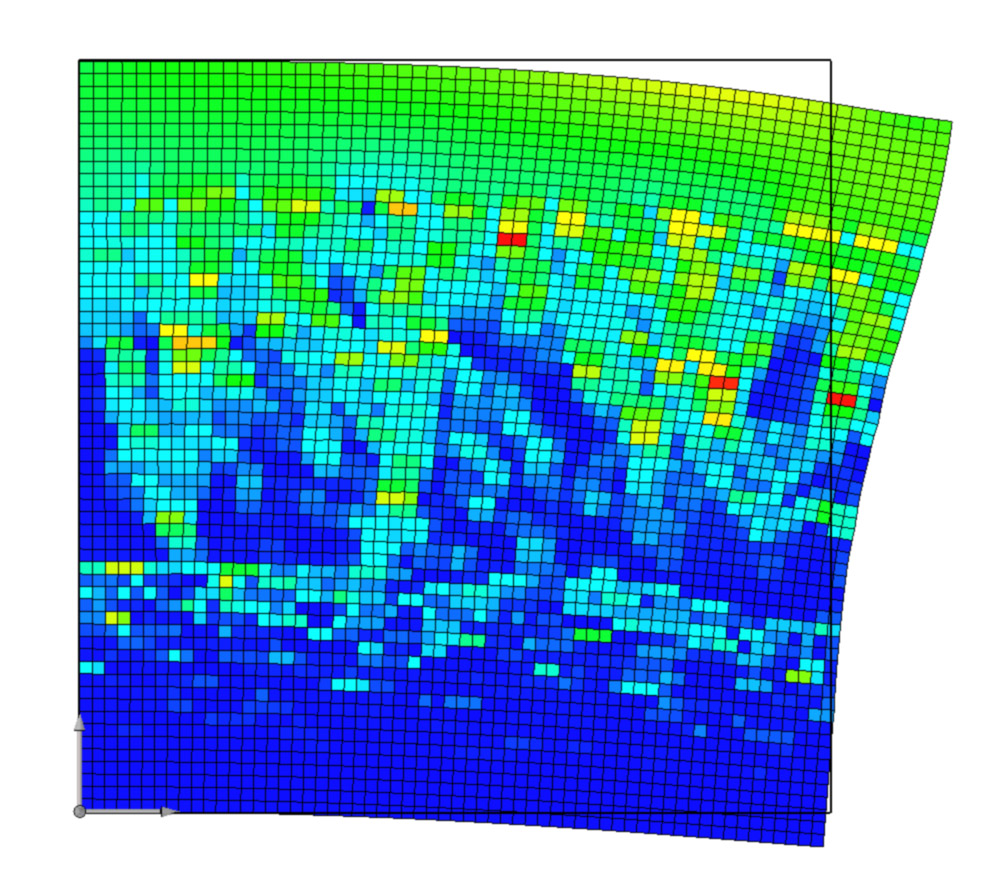}
\label{_Grad_Strain}}
\caption{Advantages of a lattice-based voxel simulation include the native ability to simulate objects with multiple interspersed materials of varying properties (a). Here, the blue material is 100 times stiffer than the red, leading to higher deformation along the top (b). Internal deformation (strain) is also shown (c).}
\label{Gradient}
\end{figure}

Limiting the discretized elements in a simulation to voxels has a number of favorable advantages. This approach enables efficient computation of the force of each constituent element, since they begin on a principle axis with identical lengths. Additionally, the stiffness of each linking beam can be pre-computed based on the stiffnesses of each constituent voxel so that each individual voxel can have a unique stiffness without altering the efficiency of the simulator (Figure \ref{Gradient}). This allows heterogeneous materials to be simulated with the same computational complexity as homogeneous materials. Additionally, using a voxel lattice eliminates the possibility of ill-formed meshes \cite{Shephard1991}. However voxel lattices are at a disadvantage to freeform meshes when large regions of homogeneous material are present or very fine local details must be simulated.

\subsection{Application in Design Automation}
Because of the exponentially increasing design space enabled by multi-material additive fabrication methods, design automation will play an increasing role in the design and optimization of structures that fully take advantage of these capabilities. Most design automation algorithms depend on many, many physical evaluations. Therefore a balance must be struck between calculating accurate results while minimizing CPU cycles. In the simulation framework presented here, static and dynamic properties are quantitatively very close to the analytical solutions for simple textbook scenarios. Features such as collision detection are in place to avoid the great inaccuracy of self intersection, but are not meant to draw scientific conclusions about the interaction between two soft bodies. By carefully budgeting CPU cycles, the simulator can accurately model physical properties while not wasting undue time on negligible effects.

\section{Physics Engine}
\subsection{Heterogeneous Deformable Body Core Simulator}
A voxel-based mass-spring lattice was chosen to best simulate the dynamics of highly deformable heterogeneous materials. Several measures were incorporated to mitigate the lack of quantitative accuracy normally associated with the discrete particle-based simulations. Each lattice point was modeled with six degrees of freedom. In addition to the traditional three translational degrees of freedom, all three rotational degrees of freedom are stored and updates as part of the state. 

It follows that not only is a mass stored for each voxel, but also its rotational equivalent moment of inertia. Instead of using simple extension springs to connect adjacent points, more complex beam elements were used that resist lateral shearing and rotation in all axes in addition to extension. By setting the properties of the beam equal to the equivalent size and stiffness of the bulk material connecting two voxels, a good approximation of the aggregate bulk material behavior is obtained.

\begin{figure}[!t]
\centering
\includegraphics[width=3.0in]{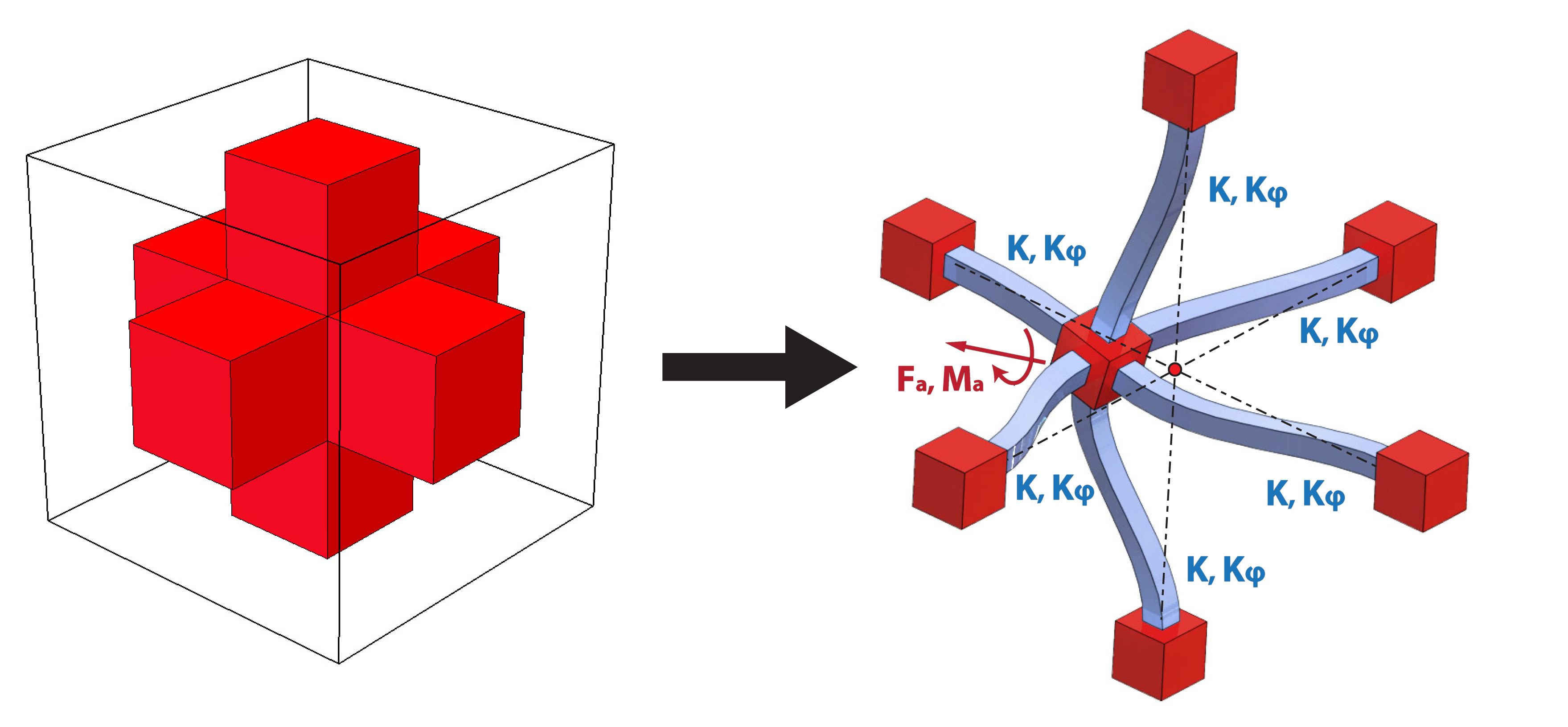}
\caption{Each voxel is modeled as a lattice point with mass and rotational inertia (red). Voxels are connected by beam elements (blue) with appropriate translational and rotational stiffnesses leading to realistic deformation under applied forces and moments.}
\label{BondVis}
\end{figure}

To prepare a given geometry for simulation, the target geometry was first voxelized into cubic voxels. Each subsequent relaxation step consists of two steps: (1) calculating all internal forces, then (2) updating all positions. In order to preserve the proper dynamics of the system positions were updated synchronously so that the order of calculation is irrelevant. In order to capture information about both translation and rotation of the voxels relative to each other, a constant cross-section beam element was used to connect adjacent voxels in the lattice. Beam elements resist both translation and rotation by exerting biaxial bending, transverse shear, and axial stretching forces in response to appropriate displacements (Figure \ref{BondVis}). Here we use a standard Bernoulli-Euler beam theory. It is important to note that the Bernoulli-Euler beam theory assumes a linearized beam model. This implies that even though the physics engine presented here is capable of modeling large aggregate non-linear deformations, the accuracy drops off as the angle between any two adjacent voxels becomes too large for a reasonable small angle approximation.

\subsubsection{Compositing Adjacent Dissimilar Materials}

The Bernoulli-Euler beam theory also requires the material to be elastic and isotropic. When two adjacent voxels are composed of the same material, the elastic modulus and stiffness of this material are used in the equivalent beam connection. However, when the materials have differing properties, an appropriate composite property must be calculated. To this end we approximate the composite stiffness of a bond between two dissimilar materials by:

\begin{equation}
E_c = \frac{2E_1E_2}{E_1+E_2}
\end{equation}

where $E_c$ is the composite elastic modulus and $E_1$ and $E_2$ are the two constituent stiffnesses. Since the elastic modulus directly corresponds to the spring constant of each bond, this is analogous to combining two springs of half the length in series with dissimilar stiffnesses. The composite shear modulus are calculated in a similar manner:

\begin{equation}
G_c = \frac{2G_1G_2}{G_1+G_2}
\end{equation}

where $G_c$ is the composite shear modulus and $G_1$ and $G_2$ are the two constituent shear moduli. Because Poisson's ratio relates the elastic modulus to shear modulus, it follows that the composite Poisson's ratio $\mu_c$ is calculated by:

\begin{equation}
\mu_c = \frac{E_c}{2G_c}-1
\end{equation}

\subsubsection{Simulation elements}

Since solid objects are represented by a network of beams connecting nodes, the physical parameters for these beams must be calculated. Because the geometry is constrained to voxels, the length $l$ of the beam was taken to be the distance between the voxels, and the cross-sectional area of the beam $A$ was $l^2$. The standard formula was used to calculate the bending moment of inertia ($I$), given that in this case both the base $b$ and height $h$ equal the lattice dimension $l$.
 
\begin{equation}
I = \frac{bh^3}{12} =  \frac{l^4}{12}
\end{equation}
 
The torsion constant ($J$) was approximated by the polar moment of inertia of a rectangular cross-section beam, and calculated as 
 
\begin{equation}
J = \frac{bh(bb+hh)}{12} =  \frac{l^4}{6}
\end{equation}
  
Using the standard Hermitian cubic shape functions for beam elements the stiffness matrix was determined for a beam element with 12 degrees of freedom: Three translational and three rotational degrees of freedom for each endpoint of the beam. This can be assembled into a stiffness matrix. The result for a beam element oriented in the positive X direction is as follows:

\begin{equation}
\begin{bmatrix}
F_{x_1} \\ F_{y_1} \\ F_{z_1} \\ M_{\theta_{x_1}} \\ M_{\theta_{y_1}} \\ M_{\theta_{z_1}} \\ F_{x_2} \\ F_{y_2} \\ F_{z_2} \\ M_{\theta_{x_2}} \\ M_{\theta_{y_2}} \\ M_{\theta_{z_2}}
\end{bmatrix}
=[K]
\begin{bmatrix}
X_1 \\ Y_1 \\ Z_1 \\ \theta_{x_1} \\ \theta_{y_1} \\ \theta_{y_1} \\X_2 \\ Y_2 \\ Z_2 \\ \theta_{x_2} \\ \theta_{y_2} \\ \theta_{z_2}
\end{bmatrix}
\end{equation}

 where the stiffness matrix $[K]$ is

\begin{equation}
[K] =
\left | 
\begin{smallmatrix} 
a_1 & 	0 & 	0 & 	0 & 	0 & 	0 & 	-a_1 & 	0 & 	0 & 	0 & 	0 & 	0 \\ 
& 		b_1 & 	0 & 	0 & 	0 & 	b_2 & 	0 & 	-b_1 & 	0 &  	0 &  	0 & 	b_2 \\ 
&	 	& 		b_1 & 	0 & 	-b_2 & 	0 & 	0 & 	0 & 	-b_1 & 	0 & 	-b_2 & 	0 \\ 
& 		&		& 		a_2 & 	0 & 	0 & 	0 & 	0 & 	0 & 	-a_2 & 	0 & 	0 \\ 
& 		& 		& 		& 		2b_3 & 	0 & 	0 & 	0 & 	b_2 & 	0 & 	b_3 & 	0 \\ 
& 		& 		& 		& 		& 		2b_3 & 	0 & 	-b_2 & 	0 & 	0 & 	0 & 	b_3\\ 
& 		& 		&		&		&		&		a_1 &	0 &  	0 & 	0 & 	0 & 	0 \\ 
&		&		&		&		&		&		&		b_1 & 	0 & 	0 & 	0 & 	-b_2\\ 
&		&		&		&		&		&		&		&		b_1 & 	0 & 	b_2 & 	0 \\ 
&		&		&		&		&		&		&		&		&		a_2 & 	0 & 	0 \\ 
&		(sym) &	&		&		&		&		&		&		&		&		2b_3& 	0 \\ 
&		&		&		&		&		&		&		&		&		&		&		2b_3
 \end{smallmatrix}
 \right |
\end{equation}

where

\begin{eqnarray}
 a_1 = \frac{E_cA}{l}\\
a_2 = \frac{G_cJ}{l}\\
b_1 = \frac{12E_cI}{l^3}\\
b_2 = \frac{6E_cI}{l^2}\\
b_3 = \frac{2E_cI}{l}
\end{eqnarray}

In order to reduce the computational complexity, the transformations to rotate each individual element from its initial orientation into the positive X direction were precomputed. Due to the cubic voxel lattice constraints, all elements are initially located on principal axes. This makes the transformation into the X-direction stiffness matrix computationally trivial. As the structure deforms over the course of a simulation, an additional transformation was calculated and applied to each element to translate the position and angle of the first voxel to zero. Therefore $X_1$, $Y_1$, $Z_1$, $\theta_{x_1}$, $\theta_{y_1}$, and $\theta_{z_1}$ all become zero and drop out of the calculation. The large matrix calculation above is reduced to:

\begin{eqnarray}
F_{x_1} = -a_1 X_2\\
F_{y_1} = -b_1 Y_2+b_2 \theta_{z_2}\\
F_{z_1} = -b_1 Z_2+b_2 \theta_{y_2}\\
M_{\theta_{x_1}} = -a_2 \theta_{x_2}\\
M_{\theta_{y_1}} = b_2 Z_2 + b_3 \theta_{y_2}\\
M_{\theta_{z_1}} = -b_2 Y_2 + b_3 \theta_{z_2}\\
F_{x_2} = -F_{x_1}\\
F_{y_2} = -F_{y_1}\\
F_{z_2} = -F_{z_1}\\
M_{\theta_{x_2}} = a_2 \theta_{x_2}\\
M_{\theta_{y_2}} = b_2 Z_2 + 2 b_3 \theta_{y_2}\\
M_{\theta_{z_2}} = -b_2 Y_2 + 2 b_3 \theta_{z_2}
\end{eqnarray}

The resulting forces and moments are then transformed back to the current orientation of the bond using the inverse of the transform calculated to arrange them in the positive X axis. The forces for all the bonds are calculated separately, then total forces ($F_t$) and moments ($M_t$) on each voxel are summed according to how many bonds $n$ are connected to it.

\begin{eqnarray}
F_t = \sum_{b=0}^{b=n}\vec{F}_b \\
M_t = \sum_{b=0}^{b=n}\vec{M}_b
\end{eqnarray}

\subsubsection{Integration}

Because momentum plays a key role in all dynamic simulations, two integrations are necessary to update the position realistically. For this physics engine, double  Euler integration was used. Although there are more accurate  integration methods, such as the Runge-Kutta (RK4) method, Euler was chosen because the massive number of discrete voxels and non-linear effects such as stick-slip friction are not well suited for the predictive steps of the RK4 integration scheme. The state of each voxel was represented by three dimensional position ($\vec{D}$) and rotation ($\vec{\theta}$) vectors and three dimensional linear ($\vec{P}$) and angular momentum ($\vec{\phi}$) vectors. In order to advance the simulation from time $t_n$ to time $t_{n+1} = t+dt$,

\begin{eqnarray}
\vec{P}_{t_{n+1}} = \vec{P}_{t_n} + F_t dt \\
\vec{D}_{t_{n+1}} = \vec{D}_{t_n} + \frac{\vec{P}_{t_{n+1}}}{m}dt \\
\vec{\phi}_{t_{n+1}} = \vec{\phi}_{t_n} + M_t dt \\
\vec{\theta}_{t_{n+1}} = \vec{\theta}_{t_n} + \frac{\vec{\phi}_{t_{n+1}}}{I}dt
\end{eqnarray}

where $m$ is the mass of the voxel and $I$ is the rotational inertia.

\subsubsection{Choosing the Timestep}
A critical aspect of implementing a robust physics simulation driven by Euler integration is to choose a suitably small timestep to prevent numerical instability. However, in order to be computationally efficient, the timestep should not be unduly small. Fortunately, it is trivial both conceptually and computationally to determine the longest stable timestep at each iteration of the simulation. In an oscillating system, the simulation will be stable if

\begin{equation} dt < \frac{1}{2 \pi \omega_{0_m}} \end{equation}
 
Because each bond between voxels is essentially a mass-spring-damper system, $\omega_{0_m}$ is simply the maximum natural frequency of any bond in the system. The stiffness of each bond was divided by the minimum mass of either voxel connected to it to calculate the maximum natural frequency of each bond according to

\begin{equation} \omega_{0_{max}} = \sqrt{\frac{k_b}{m_m}} \end{equation}
 
where $k_b$ is the stiffness of the bond and $m_m$ is the minimum of either mass connected by this bond.

\subsubsection{Damping}
Once an optimal timestep has been chosen, it is necessary to implement damping into the system to avoid the accumulation of numerical error as well as to enable realistically damped material properties. Because one application goal of this simulation involves unconstrained motion of soft bodies, damping must be included at the local interaction between voxels, not just applying a force proportional to each voxel's global velocity which would damp rigid body motion.

The local damping between adjacent voxels ensures that modal resonances at the scale of a single voxel do not accumulate. For each bond between two voxels, A force was applied to each voxel opposing the relative velocity between them. However, because rotational degrees of freedom allow this bond to be spinning, both angular and translational velocities must be correctly accomodated to make sure rigid body motion is not being damped. For each bond, first the average position, velocity, and angular velocity were calculated. Then the velocity of the second voxel relative  to the first ($\vec{V}_{2\rightarrow 1}$) is calculated according to

\begin{equation} \vec{V}_{2\rightarrow 1} = (\vec{V}_2 - \vec{V}_a)+(\vec{D}_2-\vec{D}_a)\times \vec{\omega}_a \end{equation}
 
where $\vec{V}_2$ is the velocity of the second voxel, $\vec{V}_a$ is the average velocity of two voxels, $\vec{D}_2$ is the position of the second voxel, $\vec{D}_a$ is the average position, and $\vec{\omega}_a$ is the average angular velocity. This is in effect subtracting out the rigid motion components of the relative velocity such that they are not damped. Then for each each voxel the damping force is calculated according to the standard linear damping formula:

\begin{equation} F_d =2 \zeta \sqrt{mk}V_r \end{equation}

where $F_d$ is the damping force to be applied to the voxels with mass $m$ attached to a bond with stiffness $k$ and a relative velocity $V_r$. The damping ratio $\zeta$ is normally selected to be 1, corresponding to critical damping. Likewise, angular velocities are also damped according to

\begin{equation} M_d = 2 \zeta \sqrt{Ik_{\phi}}\omega_r  \end{equation}

where $M_d$ is the damping force to be applied to the voxels with a rotational moment of inertia $I$ attached to a bond with rotational stiffness $k$ and a relative angular velocity $\omega_r$. The rotational damping ratio $\zeta$ is also normally selected to be unity. However, even though each bond is critically damped locally, the structure as a whole is still quite underdamped. So, each voxel was also variably damped in a similar manner relative to ground according to the situation at hand.

\subsection{Collisions}

\subsubsection{Gravity, floor and friction model}

In order to properly simulate freely moving soft bodies, gravity is necessary. As the force is summed on each voxel, the mass of the voxel times the acceleration of gravity was subtracted from the vertical component of force. In conjunction with gravity, a floor was implemented to for objects to rest on. Because the maximum simulation time step that can be taken is limited by the maximum stiffness between any two connected masses, the effective normal stiffness of the floor on any voxels in contact with it cannot be infinitely high. In order to keep the simulation as efficient as possible, the stiffness of each voxel contacting the floor was the stiffness of the floor in that location. Although this allows significant floor penetration in some cases, the qualitative behavior is appropriate. Potential collisions with the floor are trivial to detect by simply comparing the vertical position of each voxel to the ground plane, after accounting for the current size of the voxel.

Although a standard linear friction model would provide a relatively realistic simulation, much more interesting and realistic behavior can be observed using a Coulomb friction model. This implies that a voxel at rest with the floor will resist any motion until

\begin{equation} |F_l| > \mu_{s}F_n \end{equation}
 
where $F_l$ is the horizontal force parallel to the ground, $\mu_{s}$ is the coefficient of static friction between the voxel and the ground plane and $F_n$ is the normal force pressing this voxel into the plane of the ground. A boolean flag is set indicating to the simulation that this voxel should not move laterally, but can still move in the direction normal to ground such that it can be unweighted and then moved laterally. Once the static friction threshold has been exceeded at any given time step, the voxel is allowed to begin motion in the appropriate lateral direction by clearing the boolean static friction flag. The voxel is allowed to move in all three dimensions, but a friction force is applied opposing the lateral direction of motion according to:

\begin{equation} |F_l| = \mu_{d}F_n \end{equation}

where $\mu_{d}$ is the dynamic coefficient of friction. In order to properly detect when a voxel has stopped lateral motion, a minimum motion threshold must be set. Otherwise the voxel will never re-enter the static friction state until the velocity is less than the precision of a floating point variable. In order to detect a stopping voxel, especially one that would change direction and incorrectly bypass the effects of static friction, a voxel is artificially halted if

\begin{equation} V_l \leq \frac{F_n\mu_d dt}{m} \end{equation}

where $m$ is the mass of the voxel in question. Because the force of friction ($F_n\mu_d$) is always directly opposed to the voxels lateral velocity ($V_l$), the voxel is stopped if the projected change in velocity would change the direction of the surface velocity, which would involve the momentary stopping of the voxel. Collisions are also damped normal to the direction of contact with a user variable damping ratio ranging from zero (no damping) to 1 (critical damping).

\subsubsection{Self collision detection and handling}

Collision detection between voxels must be implemented carefully to avoid this being a bottleneck in CPU cycles. Especially in simulations with many independently moving particles, the $O(n^2)$ process of checking every particle against every other to detect collisions is prohibitively expensive in CPU cycles. In a voxel simulation such as this there are many ways we can make collision detection more efficient. Since large deformations and multiple bodies are possible, we cannot simply exclude collision detecting between voxels that are connected. However, we can immediately assume that voxels on the interior of an object my be disregarded for any collisions, assuming (as we will) that collisions are handled in such a way that overlaps cannot penetrate the outer shell. Upon import into the simulation, a list of surface voxels is precomputed, since this information will never change.

The next step is to build a list of voxel pairs that are within a collision horizon (reasonable range) of each other. Voxels on this shortlist should be compared at every timestep for potential overlap. The collision horizon was chosen to be a distance equivalent to two voxels. However, it is undesirable to watch for potential collisions between voxels that are adjacent and connected in the lattice, since the internal forces between them already resist penetration. To account for this, a list of voxels within a 3D manhattan distance of 3 in the lattice is precompiled upon import into the simulation for each voxel. Once complete, any potential collision interactions can be compared against this list in linear time to exclude the computational overhead of calculating spurious collision interactions.

Finally, the list of potential voxel collision pairs must be updated often enough that out-of-range voxels beyond the collision horizon do not have a chance to penetrate before being recognized as potential collisions. To accomplish this, a global maximum motion variable is initialized in the simulation. Each time step, the magnitude of the maximum velocity of any voxel in the simulation is added to this variable. While the maximum motion is less than half the collision horizon, it is guaranteed that no voxels can overlap into a collision. This is extremely conservative, but also computationally trivial to compute. 

\subsection{Volumetric Actuation}
For convenience, we will refer to volumetric actuation in the context of materials with a non-zero coefficient of thermal expansion (CTE) in conjunction with a changing "temperature" control variable. However, volumetric actuation may be physically achieved in a variety of ways, so there is no reason to assume that the results presented here are applicable only to temperature changes. There is also no reason that different materials within the simulation couldn't expand or contract out of sync to multiple independent control variables. This would be analogous to having multiple "temperatures" that only affect certain materials. But in the following discussion, we will refer to only a single temperature control variable.

With the soft body relaxation engine in place, such volumetric actuation is implemented by simple changing the nominal rest length between adjacent voxels when computing the elastic force between them. If the elastic force ($F_E$) between two voxels is normally calculated according to

\begin{equation} \vec{F}_E = K\left (  \vec{P}_2 - \vec{P}_1 - \vec{D}_{N_{P_1 \rightarrow P_2}} \right ) \end{equation}

to add in the effects of volumetric actuation, 

\begin{equation} \vec{D}_{N_{P_1 \rightarrow P_2}}\mid _{T=T_c} = (1+\frac{\alpha_1+\alpha_2}{2}(T_c-T_r))\vec{D}_{N {P_1 \rightarrow P_2}}\mid _{T=T_r} \end{equation}
 
where $\vec{D}_{N{P_1 \rightarrow P_2}}\mid _{T=T_c}$ is the modified rest distance based on the current temperature, $\alpha_1$ and $\alpha_2$ are the coefficients of thermal expansion of the bond's constituent materials, $T_c$ is the current temperature and  $T_r$ is the reference temperature, at which there is no temperature-based expansion or contraction.

\section{Validation}
In order to ensure that the physics engine was performing properly, we compared both static and dynamic behaviors of cantilever beams to finite element and analytical solutions. To verify the static behavior of the simulation, beam deflections of both thin and think cantilever beams were compared to a linear direct stiffness method and (in the case of the thin beam) to the analytical solution. The results are outlined in Table \ref{Cantilever_Disp_Table}. For the thin beams, 20x1x1 voxels were used with a physical size of 1mm each, for a total beam size of 20mm long by 1mm thick. A material stiffness of 1MPa was specified. The force at the end of the beam was selected to be 0.03mN so that the displacement would be small (less than a voxel-height). This ensures that small angle approximations of the analytical solution are valid. The thick beams were modeled as 10x5x5 blocks of 1mm voxels with the same stiffness. In this case 0.1N of force was applied at the free end to achieve a non-negligible displacement. 

\begin{table}[t]
\caption{Maximum displacements of thin and thick cantilever beams}
\label{Cantilever_Disp_Table}
\centering
\begin{tabular} { p{.75in} | c c c }
Geometry & Mass/spring & Direct stiffness & Analytical \\ \hline
 Thin cantilever beam & 0.822mm & 0.823mm & 0.823mm \\ 
 Thick cantilever beam & 0.538mm & 0.546mm &  N/A \\
\end{tabular}
\end{table}

\begin{figure}[!t]
\centering
\subfloat[]{
\includegraphics[width=2.5in]{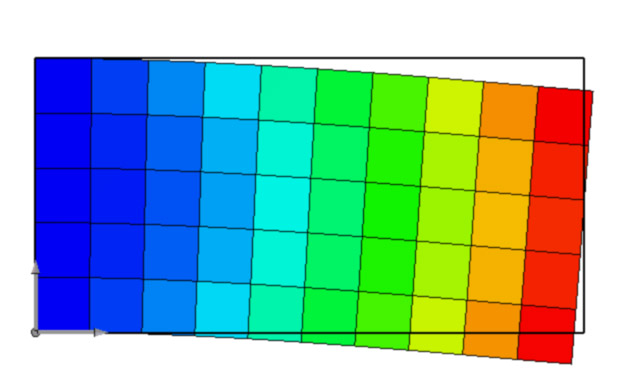}
\label{_Thick_FEA}}\
\subfloat[]{
\includegraphics[width=2.5in]{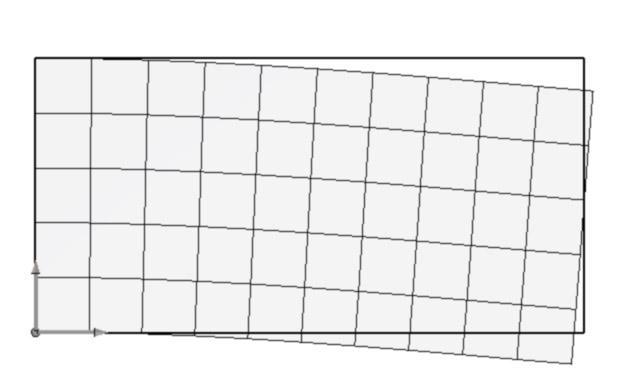}
\label{_Thick_Relax}}
\caption{The deflection of a thick cantilever beam as calculated by the direct stiffness method (finite element analysis) (a), and the mass/spring method (b).}
\label{Thick_beams}
\end{figure}

The non-linear mass/spring method presented here results in slightly smaller displacements than the direct stiffness method and the analytical solutions (Table \ref{Cantilever_Disp_Table}). This difference is negligible in the thin beam case. The difference is more pronounced in the thick beam case. This is likely because the deformation (Figure \ref{Thick_beams}) is large enough that the change in geometry in the relaxation method factors in to the results. Therefore we suspect that in this case the linear methods slightly over-predicting the deflection.

\begin{figure}[!t]
\centering
\includegraphics[width=3.0in]{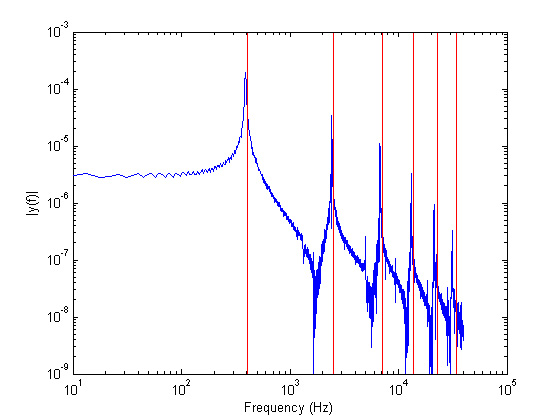}%
\caption{The frequency response of a simulated cantilever beam with low damping clearly shows modal resonances that agree well with analytically calculated values (red overlay lines).}
\label{Freq}
\end{figure}

To verify the dynamic properties of the simulation, a thin cantilever beam of the same dimensions and properties as used in the previous section was excited with an impulse force at the free end. Damping was turned off, except a trace amount of local bond rotational damping ($\zeta = 0.01$) to maintain numerical stability. 20,000 data points were collected of the z position of the voxel at the free end of the beam, corresponding to 0.13 seconds of physical time, or about 60 oscillations of the lowest-frequency fundamental mode. The frequency characteristics of this data were then plotted with the analytically calculated natural frequencies of a thin cantilever beam. (Figure \ref{Freq}) The analytical natural frequencies were calculated according to

\begin{equation} \omega_n = K_n\sqrt{\frac{EI}{\bar{m}L^4}} \end{equation}

where $\omega_n$ is the natural frequency of mode $n$ in radians per second, $K_n$ is the standard scaling factor for this mode, and $\bar{m}$ is the mass per unit length of the beam \cite{Volterra1965}. These values are overlaid on Figure \ref{Freq} and the simulated and analytical natural frequencies are tabulated in Table \ref{NatFreq}. The simulation predicts natural frequencies that are slightly lower than those predicted by beam theory. This is likely because the 20x1 aspect ratio of the simulated beam is not quite an ideal thin beam, and because there is a small amount of damping in the simulation which would tends towards under-predicting natural frequencies.

\begin{table}[t]
\caption{The modal frequencies of a simulated thin beam and analytically calculated results agree well, with the dynamic mass/spring simulation under-predicting slightly due to damping and having a finite thickness.}
\label{NatFreq}
\centering
\begin{tabular} { c | c c }
Mode & Analytical & Mass/spring \\ \hline
1 & 389Hz & 404Hz \\ 
2 & 2428Hz & 2531Hz \\ 
3 & 6749Hz & 7087Hz \\ 
4 & 13070Hz & 13890Hz \\ 
5 & 21260Hz & 22960Hz \\ 
6 & 31180Hz & 34290Hz \\ 
\end{tabular}
\end{table}

\section{Results}
\subsection{Simulation Performance}
\begin{figure}[!t]
\centering
\includegraphics[width=3.0in]{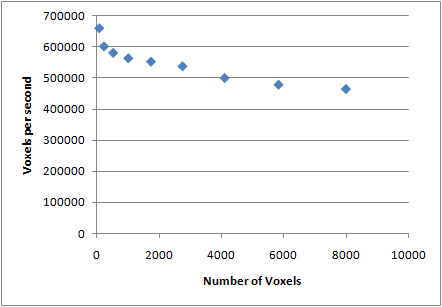}%
\caption{The computational speed per voxel drops off slightly as the number of voxels in the simulation increases.}
\label{VPS}
\end{figure}

Several parameters were explored to characterize the performance of the soft body simulator. All results presented here assume the simulation is run on a single worker thread of CoreI7 CPU at 2.67GHz. As implemented, the simulation proved very computationally efficient. For a reasonable size object of 4000 voxels, 122 complete simluation iterations were completed per second, or approximately 500,000 voxel calculations per second. As the number of voxels of increases in the object, the total voxels calculated per second decreases, but not dramatically. The simulation speed per voxel for cubic blocks of various numbers of elements are shown in Figure \ref{VPS}.

\subsection{Effects of local bond damping}
By applying a combination of damping both to the individual bonds and to the voxels relative to ground, the solution converges quickly to steady state with very little numerical jitter. The addition local damping does not significantly affect the convergence speed, but allows the solution to converge to a residual static error approximately 7 orders of magnitude lower. By suppressing jitter in this manner, both static and dynamic solutions are less susceptible to numerical instability.

\begin{figure}[!t]
\centering
\includegraphics[width=3.0in]{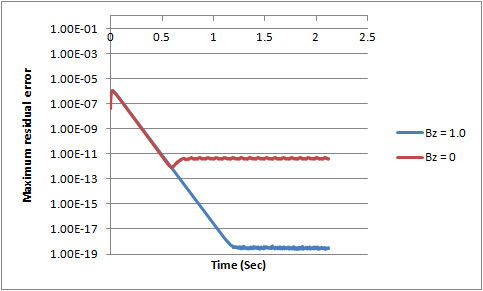}%
\caption{Damping individual bonds critically ($B_z = 1.0$) lowers the noise floor by approximately 7 orders of magnitude compared to the undamped case ($B_z = 0$).}
\label{BondDamp}
\end{figure}

\subsection{Speedup of self collision schemes}
\begin{figure}[!t]
\centering
\includegraphics[width=3.5in]{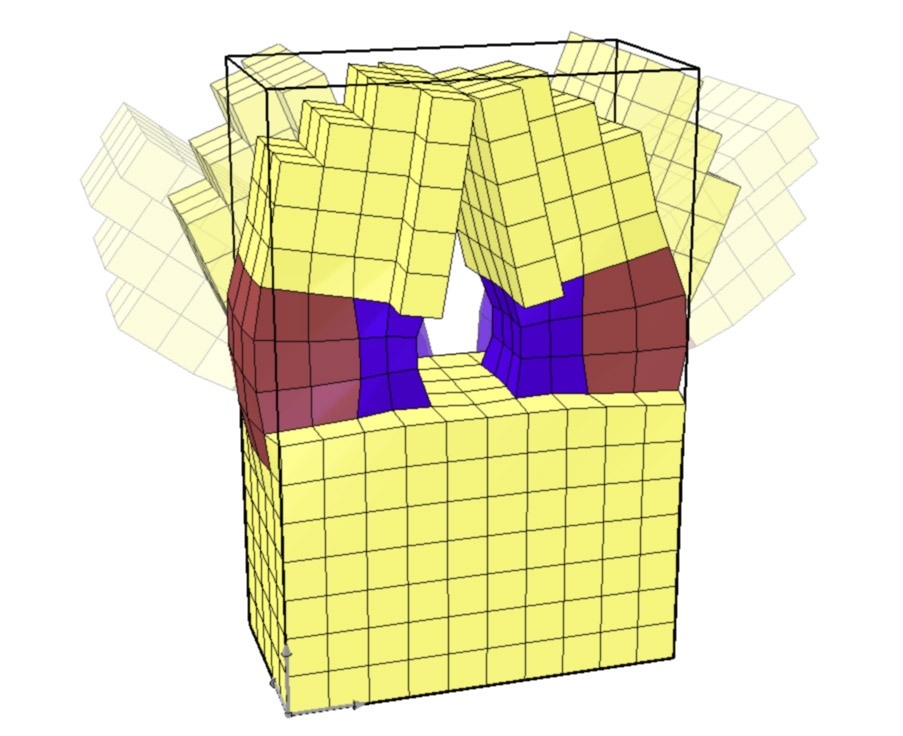}%
\caption{An arbitrary clapper setup to test net iteration rates with different collision types. The red and blue materials change volume sinusoidally 180 degrees out of phase to provide the actuation.}
\label{Clapper}
\end{figure}

Different combinations of self collision detection methods were directly compared using the test geometry shown in Figure \ref{Clapper}. All materials in this setup were defined with a stiffness of 1MPa. The red and blue materials each were assigned to have a thermal expansion coefficient with magnitude of 0.02, although one was positive and one negative. The temperature of the environment was then sinusoidally varied with an amplitude of 30 degrees, which corresponds to a 60\% expansion and contraction of the red and blue materials 180 degrees out of phase. This sets up a periodic collision between the extremities that are repeatedly entering and exiting the assigned collision horizon. 

The case of comparing the distance from all voxels to all other voxel in the structure at every timestep (All+Every) was included as a baseline. Comparing only surface voxels to each other at every timestep (Surf+Every) resulted in a very minor speedup. This is expected given the geometry chosen, because the majority of voxels in the structure are surface voxels. However, large gains in speed are realized when incorporating the collision horizon. Even when comparing all voxels to all voxels when recalculation is needed (All+Horizon) the simulation as a whole speeds up almost 6x. Again, minor acceleration is realized when considering only surface voxels with the collision horizon (Surf+Horizon).

\begin{table}[t]
\caption{Iteration rates for various collisions detection and handling schemes (Higher is better)}
\label{Collision_Table}
\centering
\begin{tabular} { l | c c c c}
Geometry & Rate (Iter/sec)\\ \hline
All+Every & 140.4\\
Surf+Every & 140.6\\
All+Horizon & 834.4\\
Surf+Horizon & 835.4\\ 

\end{tabular}
\end{table}

It should be stressed that these results are merely representative. In cases with larger numbers of voxels, the bottleneck of the entire simulation is in collision detection. In this case, comparing all voxels to all other voxels is $O(N^2)$ (where N is the number of voxels) which dominates the $O(N)$ scaling of calculating forces and updating positions. In this case, precompiling a list of surface voxels and only using them in collision detection can reduce collision detection down to approximately $O(N^{1.5})$, although this depends on the geometry. Of course, if the geometry is thin such that all voxels are on the surface, no speedup will be observed.

Additionally, the incorporating the collision horizon has extremely variable speedup up the simulation. In the extreme case of voxels moving very fast, no speedup is observed, since the collision horizon may be exceeded every timestep. However, this is only possible in extremely fast rigid body motion collisions, for which this simulation is not intended. On the opposite end of the spectrum, if the object is stationary or moving slowly, the collision horizon will take many, many timesteps to be exceeded, so almost no collision detection calculations will be needed, resulting in dramatic acceleration of the simulation.

\section{Demonstrations of simple volumetrically actuated mechanisms}

\begin{figure}[!t]
\centering
\subfloat[]{
\includegraphics[width=1.0in]{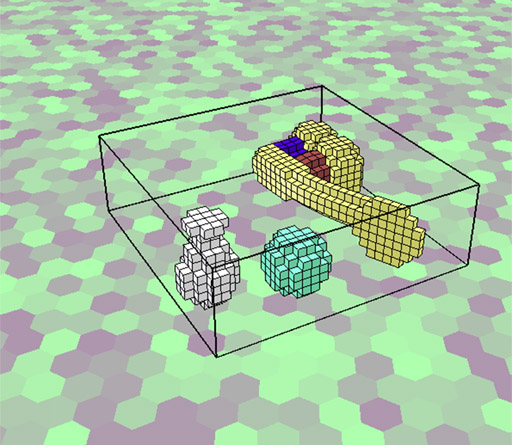}
\label{_soccer1}}
\subfloat[]{
\includegraphics[width=1.0in]{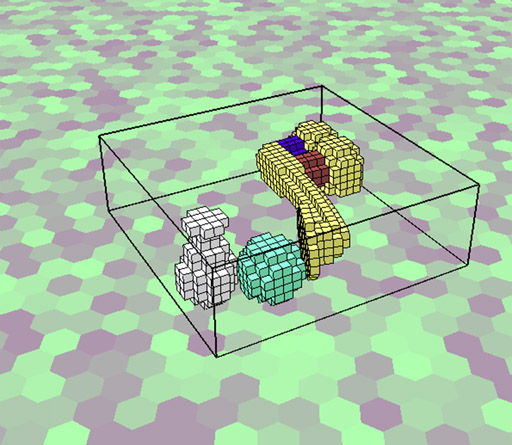}
\label{_soccer2}}
\subfloat[]{
\includegraphics[width=1.0in]{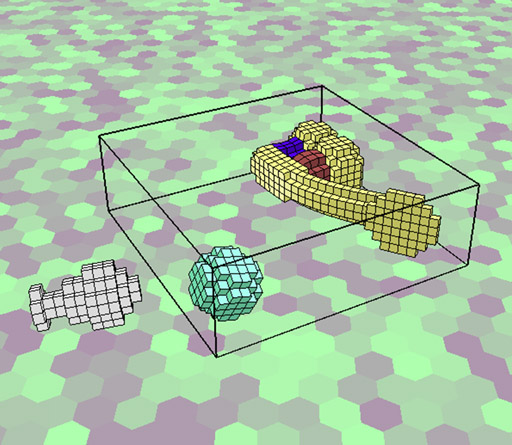}
\label{_soccer3}}\
\subfloat[]{
\includegraphics[width=1.0in]{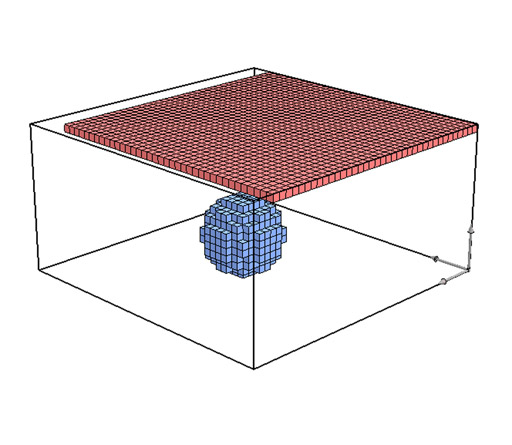}
\label{_Cloth1}}
\subfloat[]{
\includegraphics[width=1.0in]{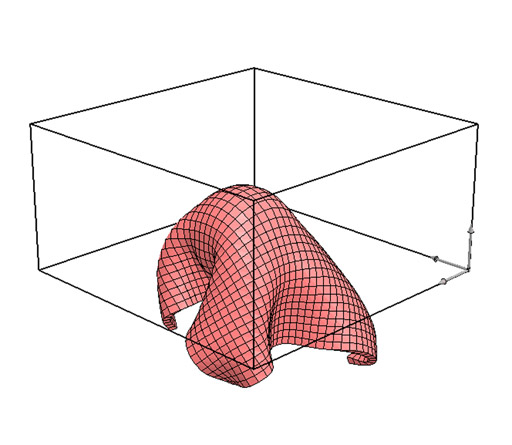}
\label{_Cloth2}}
\subfloat[]{
\includegraphics[width=1.0in]{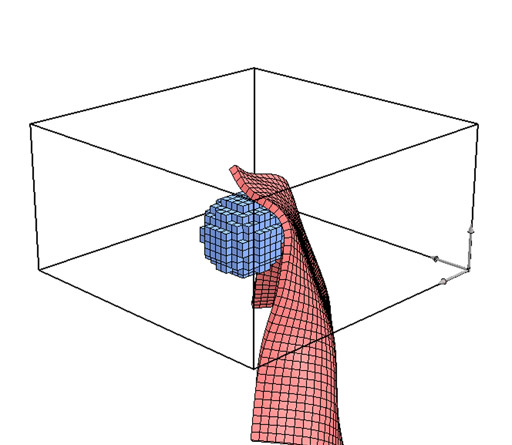}
\label{_Cloth3}}\
\subfloat[]{
\includegraphics[width=1.0in]{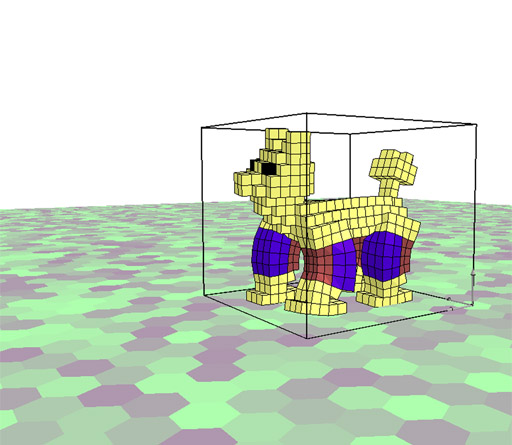}
\label{_Dog1}}
\subfloat[]{
\includegraphics[width=1.0in]{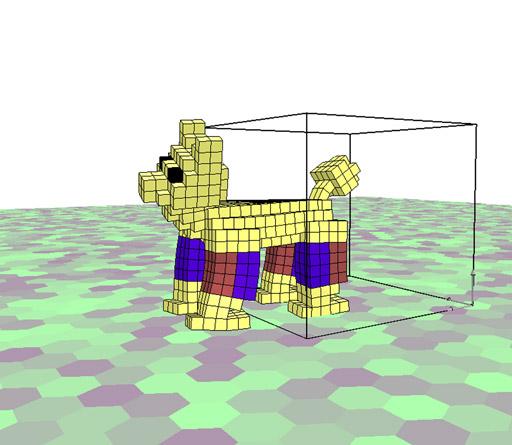}
\label{_Dog2}}
\subfloat[]{
\includegraphics[width=1.0in]{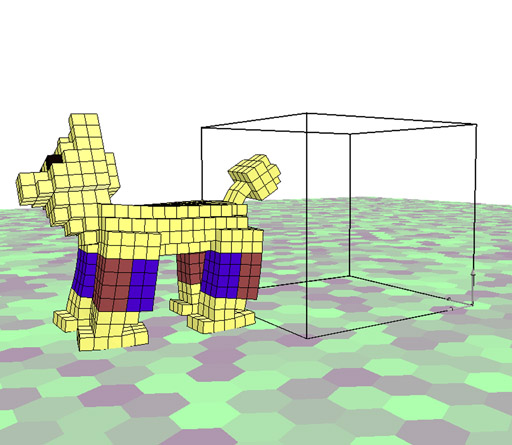}
\label{_Dog3}}\
\caption{Frames from demonstration scenes show a volumetrically actuated flexible beam swung in resonance to kick a soft ball into an even softer pin (a-c), A 2D layer of voxels falling under gravity and interacting with a fixed sphere (d-f), and a locomoting quadruped (g-i).}
\label{Soccer}
\end{figure}

Several demonstration scenes were created to illustrate the simulator in action. In the first scene, (Figure \ref{Soccer}a-c) an actuated beam kicks a ball into a bowling pin. The beam has a stiffness 10 times great than the ball, which in turn has a stiffness 10 times greater than the bowling pin. The frequency of the red and blue volumetric actuation was selected such that the beam would swing in resonance. The second scene (Figure \ref{Soccer}d-f) shows a 2D layer of voxels falling and interacting with a fixed sphere as cloth would. In the third scene, a quadruped with periodic leg actuation walks forward using the nonlinearities of the surface friction with the floor as well as the side-to-side resonance of the head swinging back and forth. These illustrations demonstrate not just the dynamics and large deformation of capabilities of the simulation, but also the use of volumetric actuation and the robust yet efficient collision system.

\section{Conclusion}

We have demonstrated a computationally efficient soft body simulator with applications in non-linear material modeling and dynamic soft object simulation. By virtue of being voxel-based, this simulation can accurately model heterogeneous materials with differing stiffnesses and densities in a physically accurate environment, even when the materials are well-interspersed among each other. This enables modeling of gradient and composite materials. By incorporating a collision horizon that is updated only when needed, self-intersection is eliminated with low computational overhead. By implementing volumetric actuation, structures can be actuated without imposing arbitrary external forces to create self-contained mechanisms.

All code and documentation is freely available at www.voxcad.com, including a standalone GUI for editing and simulating objects in a real-time interactive environment. This simulation opens the door to the design automation of a wide variety of non-linear physical structures and mechanisms that were not possible with previous soft-body physics simulation packages.

\begin{acknowledgment}
This work was supported by a National Science Foundation Graduate Research Fellowship and NSF Creative-IT grant 0757478.
\end{acknowledgment}

\bibliographystyle{plain}
\bibliography{../CCSL_JH}

\onecolumn

\appendix

\section{Simple implementation of the simulator}
\noindent
A minimal C++ program is provided here to demonstrate the ease of coding a bare-bones dynamic simulation. A 5mmx10mmx5mm beam with a material stiffness of 10MPa is cantilevered and a 1kN force is applied to the free end.

 \lstset{ %
language=C++,                
basicstyle=\footnotesize,       
frame=single,	                
tabsize=2,	                
captionpos=b,                   
breaklines=true,                
}

 \begin{lstlisting}
CVX_Object Object; //Voxel object
CVX_Environment Environment; //Environment object
CVX_Sim Simulator; //Simulator object

Object.InitializeMatter(0.001f, 5, 10, 5); //Creates a 5x10x5 workspace of 1mm voxels
int MatIndex = Object.AddMat("Material1", 10000000.0f, 0.3f); //Adds a 10MPa material to the palette
for (int x=0; x<5; x++){
	for (int y=0; y<10; y++){ 
		for (int z=0; z<5; z++){ 
			Object.SetMat(x, y, z, MatIndex); //Sets each voxel to 10MPa material
		}
	}
}
Environment.AddObject(&Object); //Imports the object into the environment
Environment.AddFixedRegion(Vec3D(0, 0, 0), Vec3D(1.0f, 0.01f, 1.0f)); //Fixes the -Y plane
Environment.AddForcedRegion(Vec3D(0, 0.99, 0), Vec3D(1.0f, 0.01f, 1.0f), Vec3D(0, 0, -1000.0f)); //Adds a downward force of 1kN to the +Y plane
			
Simulator.Import(&Environment); //Imports the environment into the simulator
Simulator.SetSlowDampZ(0.013); //Sets the global damping ratio to an appropriate value
for(int i=0; i<400; i++)Simulator.TimeStep(); //Simulates 400 timesteps
\end{lstlisting}

\section{Selected function definitions from dynamic voxel simulation classes}
\noindent
A small subset of functions used in the voxel simulator are documented here. With only these functions a dynamic simulation may be created and run. Complete documentation of the source code is available at www.voxcad.com.

\subsection{\texttt{class CVX\_Object:}}
\noindent
Describes the geometry and materials of a voxel object.\\

\noindent
{\bf \footnotesize \texttt{void CVX\_Object::InitializeMatter (float iLattice\_Dim, int xV, int yV, int zV})}\\
{\it Initializes voxel object with the specified voxel size and default lattice. A cubic lattice is assumed at the provided inter-voxel lattice dimension.}\\[0.3ex]
{\footnotesize
\begin{tabular} {|l|l|p{5.5in}|} \hline
in & iLattice\_Dim & The base lattice dimension between adjacent voxels in meters. \\ \hline
in & xV & The number of voxels in the X dimension of the workspace. \\ \hline
in & yV & The number of voxels in the Y dimension of the workspace. \\ \hline
in & zV & The number of voxels in the Z dimension of the workspace. \\ \hline
\end{tabular}\\[2ex]
}

\noindent
{\bf \footnotesize \texttt{int CVX\_Object::AddMat (std::string Name, double EMod, double PRatio, std::string *RetMessage = NULL)}}\\
{\it Appends a material to the palette. Returns the index within the palette that this material was created at. This index may change if materials are deleted from the palette. All colors and physical properties besides elastic modulus and Poisson's ratio are set to defaults. }\\[0.3ex]
{\footnotesize
\begin{tabular} {|l|l|p{5.5in}|} \hline
in & Name & The name of the material to add. If the name is already in use a variant will be generated.  \\ \hline
in & EMod & The elatic modulus of the new material in Pascals. \\ \hline
in & PRatio & The poissons ratio of the new material.  \\ \hline
out & RetMessage & Pointer to an initialized string. Messages generated in this function will be appended to the string.  \\ \hline
\end{tabular}\\[2ex]
}

\noindent
{\bf \footnotesize \texttt{bool CVX\_Object::SetMat (int x, int y, int z, int MatIndex) [inline]}}\\
{\it Sets a single voxel to the specified material. Returns true if successful. Returns false if provided indices are outside of the workspace or the material index is not contained within the current palette. }\\[0.3ex]
{\footnotesize
\begin{tabular} {|l|l|p{5.5in}|} \hline
in & x & Integer X index of voxel location to set.  \\ \hline
in & y & Integer Y index of voxel location to set.  \\ \hline
in & z & Integer Z index of voxel location to set.  \\ \hline
in & MatIndex & Specifies the index within the material palette to set this voxel to.  \\ \hline
\end{tabular}\\[2ex]
}

\subsection{\texttt{class CVX\_Environment:}}
\noindent
Describes the physical environment and boundary conditions for the voxel object. \\

\noindent
{\bf \footnotesize \texttt{void CVX\_Environment::AddObject (CVX\_Object *pObjIn) [inline]}} \\
{\it Links a voxel object to this environment. Only one voxel object may be linked at a time.}\\[0.3ex]
{\footnotesize
\begin{tabular} {|l|l|p{5.5in}|} \hline
in & pObjIn & Pointer to an initialized voxel object to link to this simulation. \\ \hline
\end{tabular}\\[2ex]
}

\noindent
{\bf \footnotesize \texttt{void CVX\_Environment::AddFixedRegion (Vec3D Location, Vec3D Size)}} \\
{\it Adds a region of voxels to be fixed to ground. All voxels touching this region will be immobile in the simulation.}\\[0.3ex]
{\footnotesize
\begin{tabular} {|l|l|p{5.5in}|} \hline
in & Location & The corner of the region closest to the origin. Specified as a percentage (in X, Y, and Z respectively) of the overall workspace. (Location.x, Location.y, and Location.z each have a range of [0.0 to 1.0]). \\ \hline
in & Size & The size of the region. Specified as a percentage (in X, Y, and Z respectively) of the overall workspace. (Size.x, Size.y, and Size.z each have a range of [0.0 to 1.0]). \\ \hline
\end{tabular}\\[2ex]
}

\noindent
{\bf \footnotesize \texttt{void CVX\_Environment::AddForcedRegion (Vec3D Location, Vec3D Size, Vec3D Force)}} \\
{\it Applies a force to a region of voxels. All voxels touching this region will have an external force applied to them. The provided force vector will be divided equally between all voxels.}\\[0.3ex]
{\footnotesize
\begin{tabular} {|l|l|p{5.5in}|} \hline
in & Location & The corner of the region closest to the origin. Specified as a percentage (in X, Y, and Z respectively) of the overall workspace. (Location.x, Location.y, and Location.z each have a range of [0.0 to 1.0]). \\ \hline
in & Size & The size of the region. Specified as a percentage (in X, Y, and Z respectively) of the overall workspace. (Size.x, Size.y, and Size.z each have a range of [0.0 to 1.0]). \\ \hline
in & Force & The force to be distributed across this region in Newtons. The force is divided equally among all voxels in the forced region. \\ \hline
\end{tabular}\\[2ex]
}

\subsection{\texttt{class CVX\_Sim:}}
\noindent
Contains the current physical state of the voxel object and advances the simulation. \\

\noindent
{\bf \footnotesize \texttt{void CVX\_Sim::Import (CVX\_Environment *pEnvIn = NULL, std::string *RetMessage = NULL)}} \\
{\it Imports a physical environment into the simulator. The environment should have been previously initialized and linked with a single voxel object. This function sets or resets the entire simulation with the new environment. }\\[0.3ex]
{\footnotesize
\begin{tabular} {|l|l|p{5.5in}|} \hline
in & pEnvIn & A pointer to initialized CVX\_Environment to import into the simulator. \\ \hline
out & RetMessage & A pointer to initialized string. Output information from the Import function is appended to this string. \\ \hline
\end{tabular}\\[2ex]
}

\noindent
{\bf \footnotesize \texttt{void CVX\_Sim::SetSlowDampZ (double SlowDampIn) [inline]}}\\
{\it Sets the damping ratio that slows downs voxels. When this is non-zero, each voxel is damped (based on its mass and stiffness) to ground. Range is [0.0 to 1.0]. Values greater than 1.0 may cause numerical instability.}\\[2ex]
{\footnotesize
\begin{tabular} {|l|l|p{5.5in}|} \hline
out & SlowDampIn & Damping ratio for damping each voxel relative to ground. \\ \hline
\end{tabular}\\[2ex]
}

\noindent
{\bf \footnotesize \texttt{void CVX\_Sim::SetCollisionDampZ (double ColDampZIn) [inline]}}\\
{\it Sets the damping ratio for voxels in colliding state. When this is non-zero, each voxel is damped (based on its mass and stiffness) according to the relative penetration velocity. Range is [0.0 to 1.0]. Values greater than 1.0 may cause numerical instability.}\\[2ex]
{\footnotesize
\begin{tabular} {|l|l|p{5.5in}|} \hline
out & ColDampZIn & Collision damping ratio for colliding voxels. \\ \hline
\end{tabular}\\[2ex]
}

\noindent
{\bf \footnotesize \texttt{void CVX\_Sim::SetBondDampZ (double BondDampZIn) [inline]}}\\
{\it Sets the damping ratio for connected voxels. When this is non-zero, each voxel is damped (based on its mass and stiffness) according to its relative velocity to the other voxel in each bond. Range is [0.0 to 1.0]. Values greater than 1.0 may cause numerical instability.}\\[2ex]
{\footnotesize
\begin{tabular} {|l|l|p{5.5in}|} \hline
out & BondDampZIn & Damping ratio between each connected voxel. \\ \hline
\end{tabular}\\[2ex]
}

\noindent
{\bf \footnotesize \texttt{bool CVX\_Sim::TimeStep (std::string *pRetMessage = NULL)}}\\
{\it Advances the simulation one time step. Given the current state of the simulation (voxel positions and velocities) and information about the current environment, this function advances the simulation by the maximum stable timestep. Returns true if the time step was successful, false otherwise.}\\[0.3ex]
{\footnotesize
\begin{tabular} {|l|l|p{5.5in}|} \hline
out & pRetMessage & Pointer to an initialized string. Messages generated in this function will be appended to the string. \\ \hline
\end{tabular}
}

\end{document}